\documentclass[english,twoside,twocolumn,prl,superscriptaddress]{revtex4}
\usepackage[latin1]{inputenc}
\usepackage{subfigure}
\usepackage{amsmath}
\usepackage{graphicx}
\usepackage{amssymb}

\makeatletter

\makeatletter
\usepackage{bm}

\makeatother

\usepackage{babel}
\makeatother
\begin{document}

\title{Attractive electron-electron interaction induced by geometric phase
in a Bloch band}

\author{Junren Shi}

\affiliation{Institute of Physics, Chinese Academy of Sciences, Beijing 100080,
China}

\author{Qian Niu}

\affiliation{Department of Physics, University of Texas at Austin, TX
  78712, USA}

\begin{abstract}
  We investigate electron pairing in the presence of the Berry
  curvature field that ubiquitously exists in ferromagnetic metals
  with spin-orbit coupling.  We show that a sufficiently strong Berry
  curvature field on the Fermi surface can transform a repulsive
  interaction between electrons into an attractive one in the $p$-wave
  channel.  We also reveal a topological possibility for turning an
  attractive $s$-wave interaction into one in the $p$-wave channel,
  even if the Berry curvature field only exists inside the Fermi
  surface (circle). We speculate that these novel mechanism might be
  relevant to the recently discovered ferromagnetic superconductors
  such as UGe$_{2}$ and URhGe.
\end{abstract}
\maketitle
Attractive interaction between electrons (or neutral fermions) is
responsible for superconductivity (or superfluidity). In condensed
matter systems, attractive interaction is usually induced by the
boson-exchange mechanism~\cite{Carbotte1990}. Indeed, in the
celebrated Bardeen-Cooper-Schreiffer (BCS) theory, electrons develop
attractive interaction by exchanging phonons~\cite{Bardeen1957}.
Subsequent studies show that other collective excitations such as
charge density waves~\cite{Kohn1965} and spin
fluctuations~\cite{Leggett1975} can also induce attractive
interaction.  The boson-exchange mechanism, together with the concept
of Cooper-pair, is considered to be the cornerstone of the modern
theory of superconductivity, and its validity can even be extended to
the unconventional superconductors such as the high-$T_{c}$
cuprates~\cite{Bednorz1986}.

In this Letter, we show a new possibility for the occurrence of
attractive electron-electron ($e$-$e$) interaction in ferromagnetic
metals with spin-orbit coupling.  We take the ferromagnetic state as
given, and focus on the effect of the Berry curvature field which
exists ubiquitously in such materials~\cite{Zak1989,Zak1989a}.  Our
question is relevant because superconductivity has been found within
ferromagnetic phase, such as in UGe$_{2}$~\cite{Saxena2000} and
URhGe~\cite{Aoki2001}.  The Berry curvature effect on electron motion
is analogous to a magnetic field in the reciprocal space
~\cite{Chang1995,Sundaram1999}, and has been invoked to successfully
explain the anomalous Hall effect in
ferromagnets~\cite{Jungwirth2002,Fang2003,Yao2004}.  On the other
hand, unlike a magnetic field in real space, monopole sources for the
Berry curvature field can occur in the reciprocal space at band
degeneracy points.  In the vicinity of the monopoles, the Berry
curvature becomes very strong.

\begin{figure}
  \subfigure[]{\includegraphics[width=0.35\columnwidth]{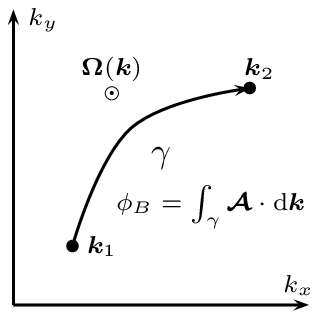}}
  \subfigure[]{\includegraphics[width=0.6\columnwidth]{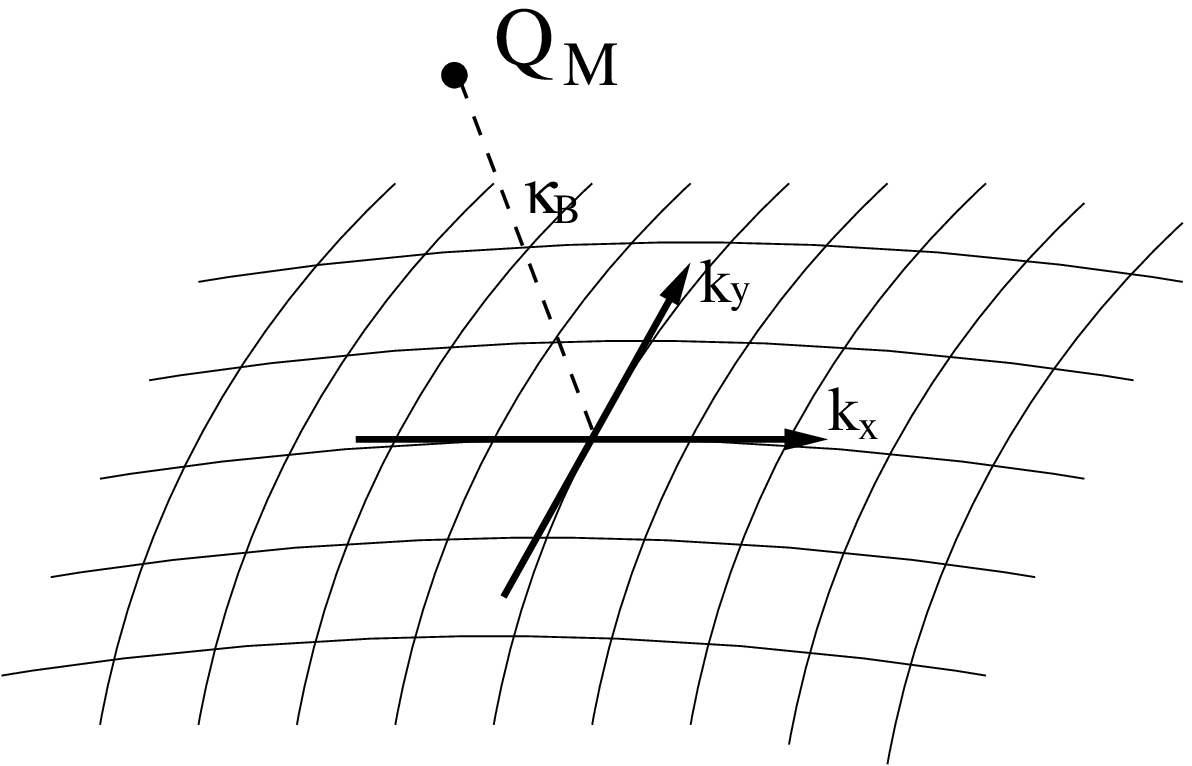}}

  \caption{\label{cap:Berry Phase} (a) Electron moving adiabatically
    in the reciprocal space acquires a Berry phase; (b) Berry
    curvature field in the vicinity of a band degeneracy split by
    magnetization and spin-orbit coupling can be modeled as a
    {}``magnetic field'' in the reciprocal space generated by a
    {}``monopole'' out of the 2D Brillouin manifold.}
\end{figure}

Our theory is formulated within an effective one band model, where
ferromagnetism and spin-orbit coupling has already been taken into
account, such as one calculated self-consistently from a spin-density
functional theory.  Figure~\ref{cap:Berry Phase}(a) shows the origin
of the Berry curvature field: an electron evolving adiabatically in
the reciprocal space will accumulate a geometric (Berry) phase
$\phi_{B}=\int_{\gamma}\langle u_{\bm k}|i\partial_{\bm k}u_{\bm
  k}\rangle\cdot\mathrm{d}\bm k$ associating with the adiabatic change
of the quasi-momentum $\bm k$~\cite{Berry1984}, in analogy to the
Aharanov-Bohm phase acquired by electron moving in the real space in
the presence of a magnetic field. It suggests a fictitious
{}``magnetic field'' in the reciprocal space with the {}``vector
potential'' $\bm{\mathcal{A}}(\bm k)=\langle u_{\bm k}|i\partial_{\bm
  k}u_{\bm k}\rangle$ and the corresponding {}``physical field''
(Berry curvature field) $\bm\Omega(\bm k)=\bm\nabla_{\bm
  k}\times\bm{\mathcal{A}}(\bm k)$, where $u_{\bm k}$ is the periodic
part of the Bloch wave function for the electron band concerned.

The central result of this work is that attractive interactions in the
$p$-wave channel may be produced with the help of the Berry curvature
field.  We show that the presence of a sufficiently strong Berry
curvature field on the Fermi surface can transform a repulsive $e$-$e$
interaction into an attractive one in the $p$-wave channel.  There is
also a topological effect analogous to the Aharanov-Bohm phase.  This
is for a situation where the Berry curvature field vanishes or is
negligible on the Fermi surface but not so inside of it.  A Berry
phase around the Fermi surface can still result from the flux within.
We show that an originally attractive interaction in the s-wave
channel can be turned into one in the $p$-wave channel.

To be specific, we investigate the following effective one-band
many-body Hamiltonian: 
\begin{equation}
  \hat{H}=\sum_{i}\epsilon(\hat{\bm k}_{i})+\sum_{i<j}V(\hat{\bm
    r}_{i}-\hat{\bm r}_{j})\,,\label{eq:effective
    Hamiltonian}
\end{equation} 
where $\epsilon(\hat{\bm k})$ is the quasi-particle dispersion
operator, and $V(\hat{\bm r}_{i}-\hat{\bm r}_{j})$ is the two-body
$e$-$e$ interaction. $\hat{\bm k}$ ($\hat{\bm r}$) is the
quasi-momentum (position) operator, and the indexes $i,\, j$ denote
the particle number. By assuming the usual canonical commutation
relations for $\hat{\bm k}$ and $\hat{\bm r}$, the Hamiltonian had
served as the basis for many successful condensed matter theories, for
instance, the conventional BCS theory.

Our study is motivated by the understanding that in the presence of
the Berry curvature field, the different components of the position
operator $\hat{\bm r}$ do not
commute~\cite{Murakami2003,Fang2003,Xiao2005}:
\begin{equation}
[\hat{r}_{\mu},\,\hat{r}_{\nu}]=i\epsilon_{\mu\nu\gamma}\Omega_{\gamma},
\label{eq:commutation relation}
\end{equation}
where $\mu,\nu=x,y,z$ denote the different components of the position
operator. It is then interesting to see how the change of the electron
dynamics dictated by the non-commutative position operator affects the
electron correlations in our system (\ref{eq:effective Hamiltonian}).

To proceed, we introduce the canonical coordinates $\hat{\bm R}$ which
satisfy the usual commutation relations
$[\hat{R}_{\mu},\,\hat{R}_{\nu}]=0$ and
$[\hat{R}_{\mu},\,\hat{k}_{\nu}]=i\delta_{\mu\nu}$, which can be
realized if we define:
\begin{equation}
  \hat{\bm R}=\hat{\bm r}-\bm{\mathcal{A}}(\hat{\bm k}),
  \label{eq:canonical coordinates}
\end{equation}
where $\bm{\mathcal{A}}({\bm k})=\langle u_{\bm k}|i\partial_{\bm
  k}u_{\bm k}\rangle $ is the {}``vector potential'' corresponding to
the Berry curvature field $\bm\Omega$.  We can then second-quantize
the many-body Hamiltonian Eq.~(\ref{eq:effective Hamiltonian}).  We
first re-express the interaction potential in terms of the canonical
coordinates $\hat{\bm R}$. This can be done by making use of the
Fourier expansion $V(\hat{\bm r}_{i}-\hat{\bm
  r}_{j})=(2\pi)^{-d}\int\mathrm{d}\bm q\upsilon(\bm q)e^{i\bm
  q\cdot\hat{\bm r}_{i}}e^{-i\bm q\cdot\hat{\bm r}_{j}}$ and the
relation $e^{i\bm q\cdot\hat{\bm r}}=e^{i\chi(\bm q,\hat{\bm
    k})}e^{i\bm q\cdot\hat{\bm R}}e^{-i\chi(\bm q,\hat{\bm k})}$,
where $\chi$ is defined by the equation $\bm q\cdot\bm\nabla_{\bm
  k}\chi(\bm q,\bm k)=\bm q\cdot\bm{\mathcal{A}}(\bm k)$.  The
interaction potential can then be expressed in the plane wave basis
$\psi_{\bm k}(\bm R)=1/\sqrt{\mathcal{V}}\exp(i\bm k\cdot\bm R)$, from
which its second-quantized form can be easily read out, yielding
finally: 
\begin{multline}
  \hat{H}=\sum_{\mathbf{k}}\epsilon(\bm k)c_{\bm k}^{\dagger}c_{\bm k}\\
  +\frac{1}{2\mathcal{V}}\sum_{\bm k\bm k^{\prime}\bm K}u(\bm k,\bm
  k^{\prime};\bm K)c_{\frac{\bm K}{2}+\bm
    k^{\prime}}^{\dagger}c_{\frac{\bm K}{2}-\bm
    k^{\prime}}^{\dagger}c_{\frac{\bm K}{2}-\bm k}c_{\frac{\bm
      K}{2}+\bm k},\label{eq:second quantization}
\end{multline} 
where $\mathcal{V}$ is the total volume of the system, $c^{\dagger}$
($c$) is the quasi-particle creation (annihilation) operator, and
\begin{equation}
  u(\bm k,\bm k^{\prime};\bm K)=\upsilon(\bm k^{\prime}-\bm k)
  e^{i\phi_{B}(\frac{\bm K}{2}+\bm k^{\prime},\frac{\bm K}{2}+\bm k)+i\phi_{B}(\frac{\bm K}{2}
    -\bm k^{\prime},\frac{\bm K}{2}-\bm k)},\label{eq:effective potential}
\end{equation}
i.e., the interaction is modified by a geometric phase defined as
$\phi_{B}(\bm k,\bm k^{\prime})=\int_{\bm k}^{\bm
  k^{\prime}}\bm{\mathcal{A}}(\bm k)\cdot\mathrm{d}\bm k$ with the
integral along the straight line connecting $\bm k$ and $\bm
k^{\prime}$ in the reciprocal space. $\phi_{B}$ is exactly the Berry
phase acquired by an electron scattered from $\bm k$ to $\bm
k^{\prime}$. In Eq.~(\ref{eq:second quantization}), we omit the spin
index, and focus on ferromagnetic systems in which the spin degrees of
freedom are fully quenched due to the strong magnetization and
spin-orbit coupling.

In the following, we investigate how the geometric phase modifies the
$e$-$e$ interaction. We notice that the distribution of the Berry
curvature field in a Brillouin zone is governed by the $\bm k$-points
of band degeneracies, which in mathematics are equivalent to
{}``magnetic monopoles'' in the reciprocal
space~\cite{Berry1984,Fang2003,Haldane2004}.  We thus focus on one of
such {}``monopoles'' and see how the geometric phase in its vicinity
modifies $e$-$e$ interaction. To identify the essential physics
without being obscured by complexities in mathematics, we limit our
study on a two dimensional (2D) ferromagnetic system.  In such a
system, there is usually no band degeneracy~\cite{Berry1984}.
However, one can usually find band near-degeneracies at high symmetry
points of the Brillouin zone which are only split by the presence of
magnetization and spin-orbit coupling. In the vicinity of these
points, the Berry curvature field can be modeled as a {}``magnetic
field'' in the reciprocal space generated by a {}``monopole'' out of
the 2D Brillouin manifold, as shown in Fig.~\ref{cap:Berry Phase}(b).
We thus have: $\bm\Omega(\bm
k)=(Q_{M}/2)\kappa_{B}/(k^{2}+\kappa_{B}^{2})^{3/2}\hat{k}_{z}$ and
\begin{equation}
  \bm{\mathcal{A}}(\bm k)=\frac{Q_{M}}{2k^{2}}
  \left(1-\frac{\kappa_{B}}{\sqrt{k^{2}+\kappa_{B}^{2}}}
  \right)\bm k\times\hat{k}_{z},\label{eq:varphi}
\end{equation}
where we assume that the {}``monopole'' is located at
$(0,0,\kappa_{B})$.  $Q_{M}=\pm1$ is the charge of the {}``monopole''.
$\kappa_{B}$ measures how close the 2D system is to the band
degeneracy. It is related to the magnitude of the band gap induced by
the magnetization and spin-orbit coupling: the larger band gap, the
larger $\kappa_{B}$ (for instance, see Ref.~\cite{Culcer2003}).

\textit{Attractive $e$-$e$ interaction induced by the Berry curvature
  field}: First, we demonstrate that the strong Berry curvature field
in the vicinity of a reciprocal space {}``magnetic monopole'' could
transform a repulsive $e$-$e$ interaction to an attractive one in
$p$-wave channel. The $e$-$e$ interaction in a typical metal can be
modeled as $V(\bm
r)=V_{0}\exp[-\kappa_{TF}(\sqrt{r^{2}+a^{2}}-a)]/\sqrt{(r/a)^{2}+1}$,
which is screened at large distances for $r\gg1/\kappa_{TF}$ and
saturates at small distances for $r\ll a$, where $\kappa_{TF}$ is the
Thomas-Fermi screening wave vector~\cite{Ando1982}. Its Fourier
transformation reads, 
\begin{equation} 
  \upsilon(\bm q)=\upsilon_{0}\frac{\exp
    \left[-a\left(\sqrt{q^{2}+\kappa_{TF}^{2}}-\kappa_{TF}\right)\right]}
  {\sqrt{(q/\kappa_{TF})^{2}+1}}\,,
  \label{eq:Interaction k space}
\end{equation} 
with $\upsilon_{0}=2\pi V_{0}a/\kappa_{TF}$.
Equation~(\ref{eq:Interaction k space}) should be considered as the
renormalized interaction between {}``dressed'' electrons resulting
from a complete treatment of a bare many-body
Hamiltonian~\cite{Anderson1961}.

For the isotropic model considered here, the effective interaction
between a pair of electrons with the opposite momentum ($\bm K=0$) can
be classified by their relative angular momentum $L_{z}=m\hbar$.  For
channel $m$, the effective potential is~\cite{Anderson1961}:
\begin{equation}
  u_{m}(k,k^{\prime})=\frac{1}{2\pi}\int_{0}^{2\pi}
  d\theta\upsilon(\bm k^{\prime}-\bm k)
  e^{2i\phi_{B}(\bm k^{\prime},\bm k)}e^{im\theta},
\label{eq:channel effective potential}
\end{equation}
where $\theta$ is the angle from $\bm k$ to $\bm k^{\prime}$ and we
have made use of the relation $\phi_{B}(-\bm k^{\prime},-\bm
k)=\phi_{B}(\bm k^{\prime},\bm k)$.

\begin{figure}[t]
\includegraphics[clip,width=1\columnwidth]{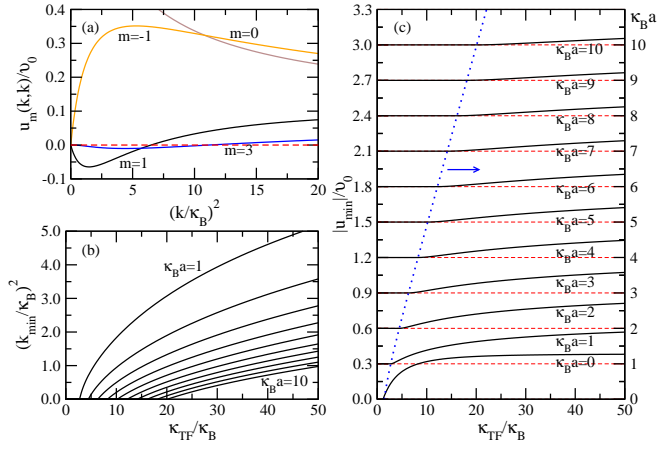}

\caption{\label{cap:Effective Potential}Effective $e$-$e$ interaction
  $u_{m}$.  (a) Typical behavior of $u_{m}(k,k)$ for different $m$.
  Parameters: $\kappa_{B}a=0.2$ and $\kappa_{TF}/\kappa_{B}=2.6$; (b)
  $k$-position of the $u_{m=1}(k,k)$ minimum as a function of
  $\kappa_{TF}/\kappa_{B}$ for different values of $\kappa_{B}a$ from
  $1$ to $10$; (c) The minimum (most attractive) value of the
  effective potential for $p$-wave channel ($m=1$) as a function of
  $\kappa_{TF}/\kappa_{B}$ for different values of $\kappa_{B}a$. The
  different curves are offset vertically for clarity. The dotted line
  shows the boundary determined from Eq.~(\ref{eq:Criteria}) for the
  onset of attractive interaction. $Q_{M}=1$.}
\end{figure}

Figure~\ref{cap:Effective Potential}(a) shows the the effective
interaction for different channels for a given set of parameters.
Attractive interaction (i.e., $u_{m}(k,k)<0$) is evident for channel
$m=1$ ($p$-wave) and $m=3$ ($f$-wave). Unlike the conventional
attractive interaction due to boson exchange which is always present
in a thin shell near the Fermi surface~\cite{Carbotte1990}, the
effective interaction due to the geometric phase is attractive only in
the vicinity of the {}``monopole'' (i.e., $\bm k=0$).
Figure.~\ref{cap:Effective Potential}(b-c) show the dependence of the
effective interaction ($k$-position where the effective interaction is
the most attractive, and its magnitude, respectively) on the
parameters ($\kappa_{B}$, $\kappa_{TF}$, $a$) for the $p$-wave channel
($m=1$). We note that the attractive interaction only occurs in
certain regime of the parameter space.

The condition for the onset of attractive effective interaction can be
determined by examining the limit of
$k,k^{\prime}\ll\kappa_{B},\kappa_{TF}$, where $\upsilon(\bm
q)\approx\upsilon_{0}\exp\left[-q^{2}/2\kappa_{u}^{2}\right]$ with
$\kappa_{u}\equiv\kappa_{TF}/\sqrt{1+\kappa_{TF}a}$ and the Berry
curvature field can be considered as a constant $\Omega_{z}(\bm
k)\approx\Omega_{0}\equiv Q_{M}/2\kappa_{B}^{2}$, with the
corresponding geometric phase: 
\begin{equation} 
\phi_{B}(\bm k,\bm
  k^{\prime})=\frac{1}{2}\Omega_{0}kk^{\prime}\sin\theta.\label{eq:Berry Phase}
\end{equation} 
Then it follows that \begin{multline}
  u_{m}(k,k^{\prime})\approx\upsilon_{0}\left|\frac{1
      -\phi_{\Omega}}{1+\phi_{\Omega}}\right|^{m/2}\exp
  \left[-\frac{k^{2}+k^{\prime2}}{2\kappa_{u}^{2}}\right]\times\\
  \left\{ \begin{array}{ll}
      I_{m}\left(\frac{kk^{\prime}}{\kappa_{u}^{2}}
        \sqrt{1-\phi_{\Omega}^{2}}\right), & |\phi_{\Omega}|\leq1\\
      (-1)^{m}J_{m}\left(\frac{kk^{\prime}}
        {\kappa_{u}^{2}}\mathrm{sgn}(\phi_{\Omega})\sqrt{\phi_{\Omega}^{2}-1}\right),
      &
      |\phi_{\Omega}|>1\end{array}\right.,\label{eq:um}
\end{multline}
where $\phi_{\Omega}\equiv\Omega_{0}\kappa_{u}^{2}$. $J_{m}$ ($I_{m}$)
is the (modified) Bessel function of the first kind. The attractive
interaction (i.e., $u_{m}(k,k)<0$) arises in the channels with odd
positive (negative) $m$ for $\Omega_{0}>0$ ($\Omega_{0}<0$) if
\begin{equation}
  |\Omega_{0}|\kappa_{u}^{2}>1\,.\label{eq:Criteria}
\end{equation}
The boundary determined from this criteria is shown as the dotted line
in Fig.~\ref{cap:Effective Potential}(c), which coincides well with
the boundary directly determined from the numerical result.

\begin{figure}[t]
\includegraphics[width=1\columnwidth]{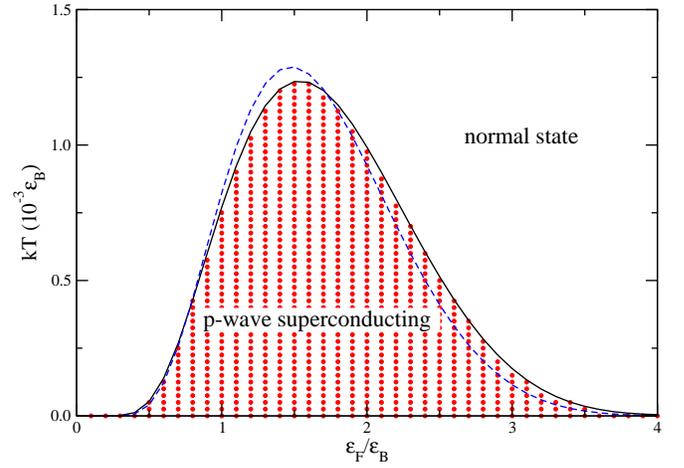}

\caption{\label{cap:phase diagram}Phase diagram in $\epsilon_{F}$
  (Fermi energy) --$T$ (Temperature) plane.
  $\epsilon_{B}\equiv\epsilon(\kappa_{B})$.  Filled dots indicate the
  region occupied by the superconducting phase.  Solid line shows the
  superconducting gap $\Delta_{F}(0)$ at the Fermi surface and at the
  zero temperature, scaled by a factor $1/1.76$.  The good
  correspondence between the phase boundary and the solid line
  suggests the usual BCS relation $\Delta_{F}(0)/kT_{c}\approx1.76$.
  The dashed line shows an empirical fitting $\Delta_{F}(0)\approx
  a\epsilon_{B}\exp[-1/\rho_{F}|u_{m}(k_{F},k_{F})|]$, where
  $\rho_{F}$ is the density of states at Fermi surface and
  $a\approx5$.  The reasonably good fitting suggests the correlation
  between the magnitude of the superconducting gap and the strength of
  the attractive interaction at the Fermi surface. Parameters:
  $\kappa_{B}a=0.2$, $\kappa_{TF}/\kappa_{B}=2.6$, and
  $\rho_{F}\upsilon_{0}=2$. The electron dispersion is assumed to be
  the simple parabolic form. $Q_{M}=1$.}
\end{figure}

We apply the BCS gap equation in Ref.~\cite{Anderson1961} to
investigate the superconducting phase induced by the attractive
interaction. The result is summarized in Fig.~\ref{cap:phase diagram}.
The superconducting state has $p$-wave symmetry with $\Delta(\bm
k)=\Delta(k)(\hat{k}_{x}\pm i\hat{k}_{y})$ (for $Q_{M}=\pm1$). The
magnitude of the superconducting gap strongly depends on the position
of the Fermi surface, which is a direct result of the strong
$k$-dependence of the effective potential.

The superconductivity we predict is closely associated with
ferromagnetism, which breaks the time-reversal symmetry, and together
with spin-orbit coupling, gives rise to the Berry curvature field in
the vicinity of the high symmetry $\bm k$-points. In this picture, the
superconducting phase naturally coexists with ferromagnetism and
disappears when the ferromagnetism is suppressed. This behavior makes
it a plausible alternative theory for the recently discovered
ferromagnetic superconductors UGe$_{2}$~\cite{Saxena2000} and
URhGe~\cite{Aoki2001}. In the traditional picture, enhanced spin
fluctuations near a quantum critical point are responsible for the
pairing of electrons. It predicts the superconducting phase on both
sides of the ferromagnetic-paramagnetic transition
point~\cite{Fay1980}, which contradicts with the experimental finding
that the superconducting phase only exists in the ferromagnetic side.
On the other hand, we note that the conditions for the onset of
superconductivity with the mechanism (i.e., Eq.~(\ref{eq:Criteria})
and the Fermi surface must reside in the vicinity of the
{}``monopole'') is rather stringent . Further investigations is
required for establishing the definite connection between the theory
and real systems.

\textit{Unconventional pairing symmetry of the topological origin}:
Second, we consider the case that the Fermi-surface (-circle) is far
from the {}``monopole'', i.e., $k_{F}\gg\kappa_{B}$. While there is no
strong presence of the Berry curvature field on Fermi-circle in this
case, the {}``monopole'' still presents a reciprocal space
{}``magnetic flux'' threading through the Fermi-disc with a total flux
$\Phi_{B}=\pi Q_{M}$. The corresponding geometric phase in the
vicinity of the Fermi-circle reads:\begin{equation} \phi_{B}(\bm k,\bm
  k^{\prime})\approx\frac{Q_{M}}{2}\theta.\end{equation} Using
Eq.~(\ref{eq:channel effective potential}), the effective $e$-$e$
interaction in channel $m$ is:\begin{equation}
  u_{m}(k,k^{\prime})\approx\upsilon_{m+Q_{M}}(k,k^{\prime})\,,\end{equation}
where $\upsilon_{m}$ is the Fourier component of the bare $e$-$e$
interaction at channel $m$. For an originally attractive interaction
in $s$-wave channel (i.e., $\upsilon_{m=0}<0)$, the effective
interaction $u_{m}$ is attractive in channel $m=-Q_{M}=\mp1$, giving
rise the $p$-wave pairing symmetry. The unconventional pairing
symmetry is of the purely topological origin: were the {}``magnetic
flux'' not present, the bare interaction would favor the $s$-wave
pairing symmetry.

The mechanism presents a new possibility distinctly different from the
conventional scenario in which the $e$-$e$ interaction induced by the
Boson-exchange mechanism has the particular structures (e.g., hot
spots induced by Fermi-surface nesting) that favors the unconventional
pairing symmetry. In contrast, the unconventional pairing symmetry
here is a result of the intrinsic structure of the Bloch band,
originated from the {}``monopole'' (band degeneracy) buried deep
inside the Fermi-sea.

So far, our discussion is built upon the Berry curvature field
associating with the adiabatic evolution of quasi-electrons in
reciprocal space.  The approach makes possible to construct a minimum
theory for the effects of the geometric phase in a Bloch band to
electron correlations.  On the other hand, it relies on the assumption
that the interaction potential varies slowly over the atomic length
scale. When the potential is not slowly varying, we have to work with
the matrix elements between the Bloch states, $\psi_{\bm k}(\bm
r)=e^{i\bm k\cdot\bm r}|u_{\bm k}(\bm r)\rangle$.  Ignoring inter-band
terms and Umklapp contributions, as usually justified, we obtain an
effective one-band Hamiltonian just like Eq.~(\ref{eq:second
  quantization}) but with (for $\bm K=0$) 
\begin{equation} 
  u(\bm k,\bm k^{\prime};\bm K=0)=\upsilon(\bm k^{\prime}-\bm k)\langle u_{\bm
    k^{\prime}}|u_{\bm k}\rangle\langle u_{-\bm k^{\prime}}|u_{-\bm
    k}\rangle.\label{eq:effective potential bloch}
\end{equation} 
In this case, the extra phases are now the Pancharatnam geometric
phases~\cite{Pancharatnam1956}. For smooth potentials, only forward
scattering is important for which $\bm k^{\prime}-\bm k$ is small,
where the Pancharatnam phases reduce to the Berry phases we discussed
before~\cite{Berry1987}.

\begin{acknowledgments}
This work is supported by the {}``Bairen'' program of Chinese Academy
of Sciences, and NSFC-10604063.
\end{acknowledgments}
\bibliographystyle{apsrev}
\bibliography{/home/shi/Papers/Papers}

\end{document}